\documentclass[11pt]{llncs}
\usepackage{amssymb}
\usepackage{amsfonts}
\usepackage{amsmath}

\newcommand{\field}[1]{\mathbb{#1}}

\newcommand{\R}{\field{R}}
\newcommand{\Z}{\field{Z}}

\newcommand{\eat}[1]{}

\newcounter{ccc}

\newcommand{\countmin}{\text{\sc Count-Min}}

  \providecommand{\abs}[1]{\lvert#1\rvert}
\providecommand{\norm}[1]{\lVert{#1}\rVert}

\providecommand{\divides}{\lvert} 
\providecommand{\card}[1]{\bigl\lvert#1\bigr\rvert}

\newcommand{\est}{\textit{est}}
\newcommand{\error}{\textit{err}}

\newcommand{\countsketch}{\text{\sc Countsketch}}

\newcommand{\crprecis}{\text{\sc CR}-precis}

\newtheorem{fact}[lemma]{Fact}

\newcommand{\rank}{\mathrm{rank}}
\newcommand{\trace}{\text{\sf freq}~}
\newcommand{\freq}{\text{\sf freq}~}

\newcommand{\extn}[1]{\ensuremath{{#1}^{e}}}

\newcommand{\approxfreq}{\text{\sc ApproxFreq}}
\newcommand{\out}{\text{output}}
\newcommand{\Space}{\text{Space}}

\title{Lower bounds on frequency  estimation of  data
streams}
\author{Sumit Ganguly\thanks{This is the full  version of the paper  with the
same title in Proceedings of the Third International Computer
Science Symposium in Russia (CSR-2008).} 
}
 \institute{Indian Institute of Technology, Kanpur }

\begin{document}

 \maketitle

\begin{abstract} We consider a basic problem in the general data streaming
model, namely, to  estimate a  vector $f \in \Z^n$ that is
arbitrarily updated (i.e.,  incremented or decremented)
coordinate-wise. The estimate $\hat{f} \in \Z^n$ must satisfy
$\norm{\hat{f}-f}_{\infty}\le \epsilon\norm{f}_1 $, that is,
$\forall i ~( \abs{\hat{f}_i - f_i } \le \epsilon \norm{f}_1)$. It
 is known to have $\tilde{O}(\epsilon^{-1})$
randomized space upper bound \cite{cm:jalgo}, $\Omega(\epsilon^{-1}
\log (\epsilon n))$ space lower bound \cite{bkmt:sirocco03} and
deterministic space upper bound of $\tilde{\Omega}(\epsilon^{-2})$
bits.\footnote{The $\tilde{O}$ and $\tilde{\Omega}$ notations
suppress poly-logarithmic factors in $n, \log \epsilon^{-1},
\norm{f}_{\infty}$ and $\log \delta^{-1}$, where, $\delta$ is the
error probability (for randomized algorithm).}
 We show that  any deterministic algorithm for
this problem  requires space $\Omega(\epsilon^{-2} (\log \norm{f}_1)
)$ bits.

\end{abstract}

\section{Introduction} \label{sec:intro}
A data stream $\sigma$ over the domain $[1,n] = \{1,2, \ldots, n\}$
is modeled as a sequence of records of the form $(\textit{pos},
i,\delta v)$, where, \textit{pos} is the current sequence index,
$i\in [1,n]$ and $\delta v \in \{+1, -1\}$. Here, $\delta v =1$
signifies an insertion of an instance of $i$ and $\delta v=-1$
signifies a deletion of an instance of $i$.
 For each data item $i \in [1,n]$, its
frequency $(\freq \sigma)_i$ is defined as  $
\sum_{(\textit{pos},i,\delta v)~ \in \text{ stream }} \delta v$. The
\emph{size} of $\sigma$ is defined as $\abs{\sigma} = \max \{
\norm{\trace \sigma'}_{\infty} \mid \sigma' $ prefix of $\sigma\}$.
In this paper, we consider the \emph{general stream model}, where,
the $n$-dimensional  frequency vector $ \freq \sigma \in \Z^n$. The
data stream model of processing permits
 online computations  over the  input sequence  using sub-linear space. The
data stream computation model has proved to be a viable model for a
number of application areas, such as network monitoring, databases,
financial data processing, etc..

We consider the  problem  \approxfreq$(\epsilon)$: given a data
stream $\sigma$, return $\hat{f}$, such that $\error(\hat{f},\freq
\sigma) \le \epsilon$, where, the function  $\error$ is given by
~\eqref{eq:error}. Equivalently, the problem may be formulated as:
given $i \in [1,n]$, return $\hat{f}_i$ such that $\abs{\hat{f}_i -
(\freq \sigma)_i} \leq \epsilon \cdot\norm{\freq \sigma}_1$, where,
$\norm{f}_1 = \sum_{i\in [1,n]} \abs{f_i}$.
\begin{equation} \label{eq:error} \error(\hat{f}, f)
\stackrel{\text{def}}{=}
\frac{\norm{\hat{f}-f}_{\infty}}{\norm{f}_1} \le \epsilon \enspace .
\end{equation}
The problem  $\approxfreq(\epsilon)$ is of fundamental interest in
data streaming applications. For general streams, this problem is
known to have  a space lower bound of $\Omega(\epsilon^{-1} \log
(n\epsilon) )$ \cite{bkmt:sirocco03}, a randomized space upper bound
of $\tilde{O}(\epsilon^{-1})$ \cite{cm:jalgo}, and  a deterministic
space upper bound of $\tilde{O}(\epsilon^{-2})$  bits
\cite{gm:escape07}.  For insert-only streams (i.e., $\freq \sigma
\ge 0$), there exist deterministic algorithms that use $O(
(\epsilon^{-1})(\log (mn)))$ space
\cite{dlm:esa02,ksp:tods03,mg:frequent82}; however extensions of
these algorithms to handle  deletions in the stream are not known.
\eat{Compressed sensing \cite{crt:ieeetit06,donoho:ieeetit06}
considers a related, though different problem; the problem is to
maintain an $m$-dimensional summary  $Ax$ of an $n$-dimensional
input vector $x$, where, the  (sensing) matrix $A$ is $m \times n$
with $m \ll n$. The works of
\cite{crt:ieeetit06,donoho:ieeetit06,indyk:uup07} show that the
 $x$ can be reconstructed from $Ax$, if $x$ is
$s$-\emph{sparse} (i.e., $x$ has at most $s$ non-zero entries),
using   deterministic space $O(s \cdot \text{ polylog}(m,n))$.
However, if $x$ not $s$-sparse, then, the reconstructed vector
$\hat{x}$ may have unbounded value of $\error(\hat{x},x)$.}

\emph{Mergeability.} Data summary structures for summarizing data
streams for frequency dependent computations (e.g., approximate
frequent items, frequency moments, etc.; formally defined in
Section~\ref{sec:straut}) typically exhibit the property of
\emph{arbitrary mergeability}. If $D$ is a data structure for
processing a stream and $D_j$, $j=1,\ldots, k$ for $k$ arbitrary, be
the respective current state of the structure after processing
streams $S_j$, then, there exists a simple operation \textit{Merge}
such that $\textit{Merge}(D_1, \ldots, D_k)$ reconstructs the state
of $D$ that would be obtained by processing the union of streams
$S_j$, $j=1, 2, \ldots, k$. For randomized summaries, this might
require initial random seeds to be shared. Thus, a summary of a
distributed stream can be constructed from the summaries of the
individual streams, followed by the \textit{Merge} operation. Almost
all known data streaming structures are arbitrarily mergeable,
including, sketches \cite{ams:jcss98},
\countsketch~\cite{ccf:icalp02}, \countmin~sketches \cite{cm:jalgo},
Flajolet-Martin sketches \cite{fm:jcss85} and its variants,
$k$-set\cite{gm:pods06}, \crprecis~structure \cite{gm:escape07} and
random subset sums \cite{gkms:vldb02}. In this paper, we ask the
question, namely, when are stream summaries mergeable?

\emph{Contributions.} We  present a space lower bound of
$\Omega(\epsilon^{-2}(\log m) ) - O(\log n)$ bits for any
\emph{deterministic} \emph{uniform} algorithm $A_n$ for the problem
$\approxfreq(\epsilon)$ over input streams of size $m$ over the
domain $[1,n]$, where, $1/(24 \sqrt{n}) \le \epsilon \le 1/32$. The
uniformity is in the sense that $A_n$ must be able to solve
$\approxfreq(\epsilon)$ for all \emph{general} input streams over
the domain $[1,n]$. The lower bound implies that the
\crprecis~structure \cite{gm:escape07} is nearly space-optimal for
\approxfreq$(\epsilon)$, up to poly-logarithmic factors.\eat{ This
solves an open problem \cite{iitk:openproblems} (Problem 4).} The
uniformity requirement is essential since  there exists an algorithm
that solves \approxfreq$(\epsilon)$ for all input streams $\sigma$
with $\abs{\sigma} \le 1$ using space $O(\epsilon^{-1}
\text{polylog}(n))$ \cite{g:ccfreq07}.

We also show that for any deterministic and uniform  algorithm $A_n$
over general streams, there exists another algorithm $B_n$ such that
(a) the state of $B_n$ is arbitrarily mergeable, (b) $B_n$ uses at
most $O(\log n)$ bits of extra space than $A_n$, and, (c) for every
input stream $\sigma$, the output of $B_n$ on $\sigma$ is the same
as the output of $A_n$ on some stream $\sigma'$ such that $\freq
\sigma = \freq \sigma'$. In other words, if $A_n$ correctly solves a
given frequency dependent problem, so does $B_n$; further, the state
of $B_n$ is arbitrarily mergeable and $B_n$ uses  $O(\log n)$ bits
of extra space. This  shows that deterministic data stream summaries
for frequency dependent computation are essentially arbitrarily
mergeable.\eat{This is the first work  to address the question of
mergeability of stream summaries. Our proof techniques combine
automata theory and algebra in a novel way.}

\section{Stream Automaton} \label{sec:straut}
In this section, we define a  stream automaton and study some basic
properties.
\begin{definition}[Stream Automaton] A  stream automaton $A_n$
over the domain $[1,n]$ is a deterministic  Turing machine that uses
two tapes, namely,  a  two-way read-write work-tape and a one-way
read-only input tape. The input tape contains the input stream
$\sigma$. After processing its input, the automaton writes an
output, denoted by $\out_{A_n}(\sigma)$, on the work-tape. \qed
\end{definition}
\emph{Effective  space usage.} We say that a stream automaton uses
space $s(n,m)$ bits  if \emph{for all} input streams $\sigma$ having
$\abs{\sigma} \le m$, the number of cells (bits) on the work-tape in
use, after having processed $\sigma$, is bounded by $s(n,m)$. In
particular, this implies that for $m \ge m'$, $s(n,m) \ge s(n,m')$.
 The space function $s(n,m)$ does not count the space
required to actually write the answer on the work-tape, or to
process the $s(n,m)$ bits of the work-tape once the end of the input
tape is observed.  The proposed model of stream automata is
non-uniform over the domain size $n$, (and uniform over the stream
size parameter $m = \abs{\sigma}$), since, for each $n \geq 1$,
there is a stream automata $A_n$ for solving instances of a problem
over domain size $n$. This creates a problem in quantifying
\emph{effective space usage}, particularly, for low-space
computations, that is,  $s(n,m)= o(n\log m)$. Let $Q(A_n)$ denote
the set of states in the finite control of the automaton $A_n$. If
$\abs{Q(A_n)} \ge m2^{n}$, then, for all $m' \le m$, the automaton
can map the frequency vector isomorphically into its finite control,
and $s(n,m) = 0$. This problem is  caused by non-uniformity of the
model as a function of the domain size $n$, and can be avoided as
follows. \eat{Let $C(A_n,m)$ be the space of the possible
configurations of $A_n$ after processing a record in the sequence,
where, each configuration is a triple of the form $(q,h,w)$, where,
$q \in Q(A_n)$, $h$ is the position of the head on the work-tape and
$w$ is the content of the $s(n,m)$-bit work tape. Therefore, the
effective space usage of $A_n$ is at least $\log \abs{C(A_n,m)}$.}
We define the effective space usage of $A_n$ as
$$\Space(A_n,m) \stackrel{\text{def}}{=} s(n,m) + \log s(n,m) +
\abs{Q(A_n)} \enspace . $$ Although, the model of stream automata
does not explicitly allow queries, this can  be modeled by a stream
automaton's capability of writing   vectors as answers, whose space
is not counted towards the effective space usage. So if
$\{q_i\}_{i\in I}$ denotes the family of all queries that are
applicable for the given problem, where, $I$ is a finite index set
of size $p(n)$ then, the output of the automaton can be thought of
as the $p(n)$-dimensional vector $\out_{A_n}(\sigma)$.

A \emph{frequency dependent problem} over a data stream is
characterized by a family of  binary predicates $P_n(\hat{f}, \freq
\sigma)$, $\hat{f} \in \Z^{p(n)}$, $n \geq 1$,  called the
characteristic predicate for the domain $[1,n]$. $P_n$   defines the
acceptability (or good approximations) of the output. A stream
automaton $A_n$ \emph{solves} a problem provided, for every stream
$\sigma$, $P_n(\out_{A_n}(\sigma), \freq \sigma)$ holds.  For
example, the characteristic predicate corresponding to the problem
$\approxfreq(\epsilon)$ is  $\error (\hat{f}, f) \le \epsilon$,
where, $\hat{f} \in \Z^n$ and $\error(\cdot,\cdot)$  is defined by
\eqref{eq:error}. Examples of frequency dependent problems are
approximating frequencies and finding frequent items, approximate
quantiles, histograms, estimating frequency moments, etc..\eat{ For
instance, if $A_n$ solves $\approxfreq(\epsilon)$ over domain
$[1,n]$, then, for any input stream $\sigma$, it suffices if
$\error(\out_A(\sigma) \in \{\hat{f}\mid \error(\hat{f},\freq
\sigma)\le \epsilon\}$.}

Given stream automata $A_n$ and $B_n$, $B_n$  is said to be an
\emph{output restriction} of $A$, provided, for every stream
$\sigma$, there exists  a stream $\sigma'$ such that, $\freq \sigma
= \freq \sigma'$ and $\out_{B_n}(\sigma) =\out_{A_n}(\sigma')$. The
motivation of this definition is the following straightforward
lemma.
\begin{lemma} \label{lem:outrestr}  Let $P_n$ be the characteristic
predicate of a frequency-dependent problem over data streams  and
suppose that a stream automaton $A_n$ solves $P_n$. If $B_n$ is  an
output restriction of $A_n$, then, $B_n$ also solves $P_n$. \qed
\end{lemma}
\begin{proof} Let $\sigma$ be any input stream to $B$ and let
$\hat{f} = \out_B(\sigma)$ be the output of $B$ on $\sigma$. Since,
$B$ is an output restriction of $A$, hence, $\hat{f} =
\out_A(\sigma)$, for some stream $\sigma$. Since, $A$ solves $P$,
therefore, $(\hat{f}, \freq \sigma') \in P$. However, $\freq \sigma'
= \freq \sigma$, and therefore, $(\hat{f},\freq \sigma) \in P$.
Since, this holds for all $\sigma$, $B$ solves $P$ as well. \qed
\end{proof}
\emph{Notation.} Fix a value of the domain size  $n \geq 2$. Each
stream record of the form $(i,1)$ and $(i,-1)$ is equivalently
viewed as $e_i$ and
 and  $-e_i$ respectively, where, $e_i = [0,\ldots, 0, 1
 $ (position $i), 0 \ldots, 0]$ is the $i^{th}$ standard basis
vector of $\R^n$. A stream is thus viewed as a sequence of
elementary vectors (or its inverse). The notation $\sigma \circ
\tau$ refers to the stream obtained by concatenating the stream
$\tau$ to the end of the stream $\sigma$. In this notation, $\freq
e_i = e_i$, $\freq -e_i = -e_i$ and $\freq \sigma \circ \tau = \freq
\sigma + \freq \tau$. The \textit{inverse stream} corresponding to
$\sigma$ is denoted as $\sigma^r$ and is defined inductively as
follows: $ e_i^r = -e_i$, $-e_i^r = e_i$ and  and $(\sigma \circ
\tau)^r = \tau^r \circ \sigma^r$. The configuration of $A_n$ is
modeled as the triple $(q, h, w)$, where, $q $ is the current state
of the finite control of $A_n$, $h$ is the index of the current cell
of the work tape, and $w$ is the current contents of the work-tape.
The processing of each record by $A_n$ can be viewed as a transition
function $\oplus_{A_n}(a,v)$, where, $a$ is the current
configuration of $A_n$, and $v$ is the next stream record, that is,
one of the $e_i$'s. The transition function is written in infix form
as $a \oplus_{A_n}  v$. We assume  that $\oplus_{A_n}$ associates
from the left, that is, $a \oplus_{A_n} u_1 \circ u_2$ means $ (a
\oplus_{A_n} u_1) \oplus_{A_n} u_2$. Given a stream automaton $A_n$,
the space of possible configurations of $A_n$ is denoted by
$C(A_n)$. Let $C_m(A_n)$ denote the  subset of configurations that
are reachable from the initial state $o$ and after processing an
input stream $\sigma$ with $\abs{\sigma} = \norm{\freq
\sigma}_{\infty} \le m$. We now define two sub-classes of stream
automata.

\begin{definition} \label{def:pathindep} A stream automaton  $A_n$  is said to be \textbf{path
independent}, if for each configuration $s$ of $A_n$ and input
stream $\sigma$, $s \oplus_{A_n} \sigma$  is dependent only on
$\trace \sigma$ and $s$. A stream automaton $A_n$ is said to be
\textbf{path reversible} if for every stream $\sigma$ and
configuration $s$, $ s \oplus_{A_n} \sigma \circ \sigma^r = s $,
where, $\sigma^r$ is the \emph{inverse} stream of $\sigma$. \qed
\end{definition}
\emph{Overview of Proof.} The proof of the lower bound on the space
complexity of \approxfreq($\epsilon$) proceeds in three steps. A
subclass of path independent stream automata, called \emph{free
automata} is defined and is proved to be the class of path
independent automata whose transition function $\oplus_{A_n}$ can be
modeled as a linear mapping of $\R^n$, with input restricted to
$\Z^n$. We then derive a space lower bound for
\approxfreq$(\epsilon)$ for  free automata
(Section~\ref{sec:free2}). In the second step, we show that a path
independent automaton that solves $\approxfreq(\epsilon)$ can be
used to design a free automaton that solves
$\approxfreq(4\epsilon)$(Section~\ref{sec:pi2}). In the third step,
we prove that for any frequency-dependent problem with
characteristic predicate $P_n$ and a stream automaton $A_n$ that
solves it, there exists an output-restricted stream automaton $B_n$
that also solves $P_n$, is path-independent, and,  $\Space(B_n,m)
\le \Space(A_n,m) + O(\log n)$. This step has two parts--- the
property is first proved for the class of path-reversible automata
$A_n$ (Section \ref{sec:prpr}) and then generalized to all stream
automata (Section \ref{sec:pnr}). Combining the results of the three
steps, we obtain the lower bound.

\section{Path-independent stream automata} \label{sec:pi} In this section, we
study the properties of path independent automata. Let $A_n$ be a
path-independent  stream automaton over the domain $[1,n]$ and let
 $\oplus$ abbreviate $\oplus_{A_n}$.\eat{ Let $ C= C(A_n)$ denote the space
of all configurations of a path independent automaton $A_n$.} Define
the function $+: \Z^n \times C(A_n)\rightarrow C(A_n)$ as follows.
\begin{align*}
x+ a = a \oplus \sigma \text{ where, } \trace \sigma = x \enspace .
\end{align*}
Since $A_n$ is a path independent automaton,  the function $x+a$ is
well-defined. The kernel $M_{A_n}$ of a path independent automaton
is defined as follows. Let the initial configuration be denoted by
$o$.
\begin{align*}
M_{A_n} = \{x \in \Z^n \mid x +o = 0 +o\}
\end{align*}
The subscript $A_n$ in $M_{A_n}$ is dropped when  $A_n$ is clear
from the context.
\begin{lemma}\label{lem:kernel} The  kernel  of a  path independent
automaton  is a sub-module of $\Z^n$.
\end{lemma}
\begin{proof} Let $x \in M$. Then, $0+o = -x + x + o = -x + o$, or
$-x \in M$. If $x,y \in M$, then, $0+o = x + o = x + y+ o$, or, $x+y
\in M$. So $M$ is a sub-module of $\Z^n$. \qed
\end{proof}
The quotient set $\Z^n/M = \{x+M \mid x \in \Z^n\}$ together with
the well-defined addition operation $(x+M) + (y +M) = (x+y) + M$,
forms a module over $\Z$. \eat{It is of central  interest from the
point of view of characterizing the computations of path independent
automata.}
\begin{lemma} \label{lem:piua} Let $M$ be the kernel of a path independent
automaton $A_n$. The mapping $x+M \mapsto x+o$ is a set
 isomorphism between $\Z^n/M$ and the  set of reachable
configurations $\{x+o\mid x \in \Z^n\}$. The automaton   $A_n$ gives
the same output for each $y \in x+M$, $x\in \Z^n$.
\end{lemma}
\begin{proof}
 $y \in x+M$ iff $x-y \in M$ or $-y + x + o = o$, or, $x+o =
 y+o$. Thus, $A_n$ attains the same configuration after processing both
$x$ and $y$ and therefore $A_n$ gives the same output for both $x$
and $y$. Since,  $x+o = y+o$ iff $x-y \in M$, which implies that the
mapping $x+M \mapsto x+o$  is an isomorphism. \qed
\end{proof}
Let $\Z_m^n$ denote the subset $\{-m, \ldots, m\}^n$ of $\Z^n$.
\begin{lemma} \label{lem:numstates2} Let $A_n$ be a path independent automaton with  kernel
$M$. Then,
 $$\Space(A_n,m) \geq \lceil~ \log \abs{\{x+M \mid
x\in \Z_m^n\}}~ \rceil  \geq
 (n -\dim M)\log (2m+1).$$
\end{lemma}
\begin{proof}
The set of  distinct configurations of  $A_n$ after it has processed
a stream with frequency $x \in \Z_m^n$ is isomorphic to $ \{x+M \mid
x\in \Z_m^n\}$. The number of configurations using workspace of $s =
s(n,m)$ is at most $\abs{Q_{A_n}}\cdot s \cdot 2^s$. Therefore,
\begin{align}\label{eq:size1}
2^{\Space(A_n,m)} = \abs{Q_{A_n}} \cdot s \cdot 2^s \geq \card{\{x+M
\mid x\in \Z_m^n\}} \enspace .
\end{align}
We now obtain an upper bound on the size
 $|M \cap  \Z_m^n|$. Let $b_1, b_2,\ldots,b_r$ be a basis for $M$.
The set
 $$P_m = \{\alpha_1 b_1 + \ldots + \alpha_r b_r \mid
\abs{\alpha_i} \leq m \text{ and integral},  i=1,2, \ldots, n \}$$
defines the set of all integral points generated by $b_1, b_2,
\ldots, b_r$ with multipliers in $\{-m, \ldots, m\}$. Thus,
\begin{align} \label{eq:McapZmn}
\abs{M \cap \Z_m^n} \le   \abs{P_m} = (2m+1)^r \enspace .
\end{align}
It follows that
\begin{align*}
\card{\{x + M\mid x \in \Z_m^n\}} \geq \frac{\abs{\Z_m^n}}{|M \cap
\Z_m^n|} \geq  (2m+1)^{n-r} \enspace.
\end{align*}
Since, $r =  \dim M$, substituting in ~\eqref{eq:size1} and taking
logarithms, we have
\begin{align*}
\Space(A_n,m) \geq  \log \card{\{x+M \mid x\in \Z_m^n\}}  \geq (n-r)
\log (2m+1)  ~. ~~~~\qed
\end{align*}
\end{proof}
Lemma~\ref{lem:makepi}  shows that given a sub-module $M$,  a
path-independent automaton with a given $M$  as a kernel can be
constructed using nearly optimal space. The transition function
$(x+M) + (y+M)= (x+y) + M$ implies that the state of a path
independent automaton is arbitrarily mergeable.
\begin{lemma} \label{lem:makepi} For any sub-module $M$ of $\Z^n$, one can construct  a
path-independent automaton with kernel $M$ that uses nearly optimal
space $s(n,m) =  \log \abs{\{ x+M \mid x \in [-m\ldots m]^n\}} +
O(\log n)$ and uses $n^{O(1)}$ states in its finite control. \qed
\end{lemma}
\begin{proof}
Let $M$ be a given sub-module of $\Z^n$ with basis $b_1, \ldots,
b_r$ (say). It is sufficient  to construct a path independent
automaton whose configurations are isomorphic to $E = \Z^n/M$.
Since, $\Z^n$ is free, $\Z^n/M$ is finitely generated using any
basis of $\Z^n$. Therefore, the basic module decomposition theorem
states that
\begin{align} \label{eq:moddecomp}
\Z^n/M = \Z/(q_1) \oplus \cdots \oplus \Z/(q_r) \enspace .
\end{align}
where, $q_1 \divides q_2 \divides \cdots \divides q_r$. (Here,
$\oplus$ refers to the direct sum of modules.) The finite control of
the  automaton stores $q_1, \ldots, q_r$  and the machinery required
to calculate $1 \mod q_j$ and $-1\mod q_j$ for each $j$. For the
frequency vector $f$, the residue vector $f+M$ is maintained as a
vector of residues with respect to the $q_j$'s as given by
~\eqref{eq:moddecomp}. Since, ~\eqref{eq:moddecomp} is a direct sum,
hence, the space used by this representation is optimal and equal to
$\abs{\{x+M \mid x \in [-m \ldots m]^n\}}$. \qed
\end{proof}
\begin{definition}[Free Automaton] \label{def:torsionfree} A path independent
automaton $A_n$ with kernel $M$ is said to be free if   $\Z^n/M$ is
a free module.\qed
\end{definition}
That is, $A_n$ is free if for every $x \in \Z^n$ such that
 there exists $a \in \Z $, $a \neq 0$  and $ax \in M$, it is the case that $x
 \in M$. For free automata $A_n$, it follows that $\Z^n$ is the direct sum of
$M$ and $\Z^n/M$, that is, $\Z^n  = \Z^n/M \bigoplus M$. For the
$\approxfreq$ problem and other related problems, it will suffice to
consider only free automata\footnote{ There exist stream automata
that use finite field arithmetic and consequently have torsion, for
example \cite{gm:pods06}.}. Lemma~\ref{lem:vecspace} shows that the
transition function $\oplus$ of a free automata can be represented
as a linear mapping.
\begin{lemma}\label{lem:vecspace}
Let $A_n$ be free   automaton  with kernel   $M $. There exists a
unique vector subspace $M^e$ of $\R^n$ of the smallest dimension
containing  $M$. The mapping $x + M \mapsto x + M^e$ is an injective
mapping from $\Z^n/M$ to $\R^n/M^e$. If  $\dim \Z^n/M = r$,   then,
there exists an orthonormal basis $V = [V_1, V_2]$ of $\R^n$ such
that $\rank(V_1) = r$, $\rank(V_2) = n-r$, $\extn{M}$ is the linear
span of $V_2$ and $\R^n/M^e$ is the linear span of $V_1$. \qed
\end{lemma}
 \begin{proof} $\Z$ is a principal and entire ring. Since
$\Z^n$ is a module over $\Z$, its sub-modules are free modules.
Therefore, $M$ is a free module. Since $\Z^n/M$ is given to be free,
$\Z^n$ is the direct sum of two free modules, $\Z^n = \Z^n/M
\bigoplus M$. Therefore, both $M$ and $\Z^n/M$ have bases, say $B_1$
and $B_2$ whose  union is a basis for $\Z^n$. Since, $\Z^n$ is a
free module and has the standard $n$-dimensional basis $e_1, \ldots,
e_n$, therefore, all bases of $\Z^n$ have the same dimension.
Without loss of generality, therefore, let $B = [b_1, b_2, \ldots,
b_n]$ be a basis of $\Z^n$ such that $B_2 = [b_1, \ldots, b_r]$ is a
basis for $M$ and $B_1=[b_{r+1}, \ldots, b_{n}]$ is a basis for
$\Z^n/M$.

 Let $M^e$ denote the span of $b_1, \ldots, b_r$ over
$\R$. $M^e$ is obviously the smallest vector space over $\R$ that
contains $M$, since, every vector space over $\R$ containing $M$
must contain the span of $b_1, \ldots, b_r$. Therefore, $\dim M^e
\leq r$ and therefore, $\dim \R^n/M^e \leq n-r$ (same argument).
However, the standard basis $\{e_1, \ldots, e_n\}$ is a basis of
$\Z^n$ and therefore, $\dim M^e + \dim \R^n/M^e = n$. Hence, $\dim
M^e =r$ and $ \dim \R^n / M^e = n-r$. Further, $b_1, \ldots, b_n$
continues to be a basis for $\R^n$, of which $b_1, \ldots, b_r$ is a
basis for $M^e$ and $b_{r+1}, \ldots, b_n$ is a basis for
$\R^n/M^e$.

Consider the mapping $x + M \mapsto x + M^e$. Let $\bar{x}$,
$\bar{y}$ denote the elements $x +M$ and $y+M$ of $\Z^n/M$. Suppose
that $\bar{x} \neq \bar{y}$. Then, $x-y \not\in M$. $x-y$ can be
expressed uniquely as a linear combination of the basis elements.
\begin{align*}
x-y = \sum_{j=1}^n \alpha_i b_i, ~~\alpha_i \in \Z
\end{align*}
Hence, $x-y$ has the same unique representation in the vector space
over $\R^n$. Further, at least one of the coordinates $\alpha_1,
\ldots, \alpha_r$ is non-zero, otherwise, $x-y$ would belong to $M$.
Since, $x-y$ has the same representation in the vector space $\R^n$,
$x-y$ is not in $M^e$. The mapping $x+M \mapsto x+M^e$ is therefore
injective. Using standard Gram-Schmidt orthonormalization of $B_1$
and $B_2$ respectively viewed as defining vector sub-spaces over
$\R$, we get $V_1$ and $V_2$. By the previous argument, $\rank(V_1)
= n-r$ and
 $\rank(V_2) = r$. \qed
\end{proof}

\eat{The space  lower bound presented in Lemma~\ref{lem:numstates2}
is easily met for free automata.
\begin{lemma} \label{lem:numstates3} Given  a sub-module $M$ of $\Z^n$
 such that $\Z^n/M$ is free, there exists a  free automaton $A_n$ with  kernel
$M$ that has  $\abs{Q_{A_n}} = n^{O(1)}$ states in its finite
control, and, uses $s(n,m) = \Theta( (n -\dim M)\log m )$ bits on
its work-tape.
\end{lemma}
\begin{proof} The construction of $A_n$ is given by Lemma~\ref{lem:makepi}.
Since, $\Z^n/M$ is free, therefore, $(M^e-M) \cap \Z^n = \phi$.
Therefore, $\abs{\{x+M\mid x \in [-m\ldots m]^n\}} = \Theta(
(2m+1)^{n-\dim M})$. Using Lemma~\ref{lem:makepi}, the statement
follows. \qed
\end{proof}
} \eat{Let $A_n$  perform the following operations. Suppose $V_1$ be
an orthonormal basis of $R^n/M^e$ (where, $M^e$ is the smallest
vector space extension of $M$ into $\R^n$). $A_n$ stores $V_1V_1^T$
and maintains $V_1V_1^T x$ on its work-tape, where $x$ is the
frequency vector of the stream seen so far. For each elementary
update of the form $\pm e_i$ that appears on the  tape, $A_n$ adds
$\pm V_1V_1^T e_i$ (i.e., column $i$ of $V_1 V_1^T$). Storing $V_1
V_1^T$ requires $O(n^2)$ scalars, and, the machinery of adding the
$ith$ column of $V_1 V_1^T$  to the work-tape can be easily
implemented using an additional  $O(n^2)$ states. \qed
\end{proof}}

\section{Frequency estimation} \label{sec:freqest}
In this section, we present a space lower bound for
$\approxfreq(\epsilon)$ using  path-independent automaton. Recall
that a stream automaton $A_n$  solves $\approxfreq(\epsilon)$,
provided, after processing any input stream $\sigma$ with $\trace
\sigma = x$, $A_n$ returns a vector
  $\hat{x} \in \R^n$ satisfying
$ \error(\hat{x},x) = \frac{\norm{\hat{x} - x}_{\infty}
}{\norm{x}_1} \leq \epsilon $.
 In
general, if an estimation algorithm returns the same  estimate $u$
 for all elements of a set $S$, then,   $\error(u,
S)$ is defined as $\max_{y \in S} \error(u,y)$. Given a set $S$, let
$\min_{\ell_1}(S)$ denote the element in $S$ with the smallest
$\ell_1$ norm:
 $ \min_{\ell_1}(S) = \text{argmin}_{ y \in S}  ~~ \norm{y}_1$.
\eat{For a coset $x+M$, $\error(\min_{\ell_1}(x+M), x+M)$ is defined
as $ \max_{y \in x+M} \error(\min_{\ell_1}(x+M), y)$. }
\begin{lemma} \label{lem:l1min} If $S \subset \Z^n$ and there exists  $h \in \R^n$
such that   $\error(h, S) \leq \epsilon$, then $\error
(\min_{\ell_1}(S), S) \leq 2\epsilon$.
\end{lemma}

\begin{proof}
Let $g$ denote $\min_{\ell_1}(S)$ and  $y \in S$. Since, $\norm{g}_1
\le \norm{y}_1$, by triangle inequality,
\begin{align*}
\error (g,y ) & = \frac{\norm{g - y}_{\infty}}{\norm{y}_1} \leq
\frac{\norm{g - h}_{\infty}} {\norm{y}_1}
 + \frac{\norm{h  - y}_{\infty}}{\norm{y}_1}\eat{, ~~~~  \text{ by triangle inequality } \\
 &} \leq \frac{\norm{g - h}_{\infty}} {\norm{g}_1} +
 \frac{\norm{h  - y}_{\infty}}{\norm{y}_1} \leq \epsilon + \epsilon =
 2\epsilon~~\text{\qed} \end{align*}
\end{proof}

\subsection{Frequency estimation using free  automata}
\label{sec:free2} In this section, let $A_n$ be a free automaton
 with kernel $M$ that solves the problem
$\approxfreq(\epsilon)$.\eat{ As shown in Section~\ref{sec:straut},
streams with frequency vector from the same coset of $M$ map to the
same configuration of $A_n$, and hence, yield the same output.
Suppose the vector output by $A_n$ when in the configuration
corresponding to the coset $x+M$ is $h(x+M)$.}
\begin{lemma}  \label{lem:zero}  Let $M$ be a sub-module of $\Z^n$.  (1) if there exists $h$ such that
$\error(h, M) \leq \epsilon$, then, $\error(0,M) \leq \epsilon$,
and,
 \label{lem:errorMe} (2)  if $\error(0,M) \leq \epsilon$ then $\error(0,M^e) \leq \epsilon$.
\end{lemma}

\begin{proof} [of Lemma~\ref{lem:zero}part (1)]For any $y_i \in \Z$, $\max(\abs{h_i-y_i}, \abs{h_i+y_i})$ $\geq \abs{y_i}$.
 Therefore, $$\max(\norm{h-y}_{\infty},\norm{h+y}_{\infty}) \geq \norm{y}_{\infty}\enspace .$$
Let $y \in M$. Since, $M$ is a module, $-y \in M$. Thus,
\begin{align*}
\error(0,y) & = \error(0,-y) = \frac{\norm{y}_{\infty}}{\norm{y}_1}
\leq \frac{1}{\norm{y}_1}
\max(\norm{h-y}_{\infty},\norm{h+y}_{\infty}) \\
& = \max(\error(h,y), \error(h,-y)) \leq \epsilon
  & \text{ \qed }
\end{align*}
\end{proof}

\begin{proof}[of Lemma~\ref{lem:errorMe} part (2)] Let $z \in M^e$.
Let $b_1, b_2, \ldots, b_r$ be a basis of the free module $M$. For
$t>0$, let $tz$ be expressed uniquely as $tz =  \alpha_1 b_1 +
\ldots + \alpha_r b_r$, where, $\alpha_i$'s belong to $\R$. Consider
the vertices of the parallelopiped $P_{tz}$ whose sides are $b_1,
b_2, \ldots, b_r$ and that encloses $tz$.
\begin{multline*}
P_{tz} = [\alpha_1]b_1 + [\alpha_2]b_2 + \ldots + [\alpha_n]b_n \\
 +\{ \beta_1 b_1 + \beta_2 b_2 + \ldots + \beta_r b_r\mid \beta_j
 \in \{0,1\}, j=1,2, \ldots, r \}
\end{multline*}
where, $[\alpha]$ denotes the largest integer smaller than or equal
to $\alpha$. Since, $\ell_{\infty}$ is a convex function $
\norm{tz}_{\infty} \leq \norm{y}_{\infty}$ for some $y \in P_{tz}$.
Let $y = \sum_{j=1}^r \beta_j b_j$, for  $\beta_j \in \{0,1\}$,
$j=1,2, \ldots, r$. \begin{align*} \norm{y- tz}_1  & = \norm{
\sum_{j=1}^r (\beta_j - [\alpha_j]) b_j }_1 \leq \sum_{j=1}^r
\norm{(\beta_j - [\alpha_j]) b_j }_1 \leq \sum_{j=1}^r
\norm{b_j}_1\\  ~~~ \text{ or, } &~~~ \norm{tz}_1 \geq \norm{y}_1 -
\sum_{j=1}^r \norm{b_j}_1
\end{align*}
Therefore, \begin{align*} \error(0,tz) &=
\frac{\norm{tz}_{\infty}}{\norm{tz}_1} \leq \frac{\norm{y}_{\infty}}
{ \norm{y}_1 - \sum_{j=1}^r \norm{b_j}_1}
\\ &\leq\left(\frac{\norm{y}_1}{\norm{y}_{\infty} } -
\frac{\sum_{j=1}^r \norm{b_j}_1}{\norm{y}_{\infty}}\right)^{-1} \leq
\left(\frac{1}{\epsilon } - \frac{\sum_{j=1}^r
\norm{b_j}_1}{\norm{y}_{\infty}}\right)^{-1}
\end{align*}
where, the last step follows from the assumption that $y \in M$ and
therefore, $\error(0,y) = \frac{\norm{y}_{\infty}}{\norm{y}_1} \leq
\epsilon$. The ratio $ \frac{\sum_{j=1}^r
\norm{b_j}_1}{\norm{y}_{\infty}}$ can be made arbitrarily small by
choosing $t$ to be arbitrarily large. Thus, $\lim_{t \rightarrow
\infty} \error(0,tz) \leq  \epsilon$. Since, $\error(0,tz) =
\frac{\norm{tz}_{\infty}}{\norm{tz}_1} =
\frac{\norm{z}_{\infty}}{\norm{z}_1} =\error(0,z)$, for all $t$, we
have, $\error(0,z) \leq \epsilon$. \qed
\end{proof}

\eat{\begin{lemma} \label{lem:errorMe} Let $M$ be a sub-module of
$\Z^n$. If $\error(0,M) \leq \epsilon$ then $\error(0,M^e) \leq
\epsilon$.
\end{lemma}}
\eat{For a  free automaton $A_n$ with kernel $M$ and corresponding
$n \times r$ orthonormal  matrix $V_1$ whose columns are a basis for
$\R^n/M^e$ (as given by Lemma~\ref{lem:vecspace}), the minimum
$\ell_2$ estimator is defined as
\begin{align} \label{lem:est2}
\bar{x}_2 = \est_2(x) =  V_1V_1^T x \enspace.
\end{align}
It is easy to show that the $\ell_2$ estimator is well-defined.
\eat{ To show that the $\ell_2$ estimator is well-defined, let
$x+M^e = y+M^e$, or that, $x-y \in M^e$. Since, the columns of $V_2$
form a basis for $M^e$, $x-y = V_2 z$, for some $z$. Therefore, $
V_1V_1^T(x-y) = V_1V_1^TV_2 z = V_1 \cdot 0 \cdot z = 0 $, or  that
$V_1V_1^Tx = V_1V_1^Ty$, or equivalently, $\bar{x}_2 = \bar{y}_2$.}
It is called the minimum $\ell_2$ estimator since it returns a point
in the coset $x+M^e$ that is closest to the origin in terms of the
$\ell_2$ distance.}\eat{ Equivalently, $\est_2(x)$ could be defined
as the element in $x+M^e$ with the smallest $\ell_2$ norm:
$$\est_2(x) = \text{argmin}_{y \in x+M^e} \norm{y}_2 \enspace . $$
 We  now  show that there is a subset $J$ of the set of the
standard unit vectors $\{e_1, e_2, \ldots, e_n\}$  such that
$\abs{J} = \Theta(n)$ and the minimum $\ell_2$ estimator is nearly
optimal for the unit vectors in $J$}
\begin{lemma} \label{lem:J} Let $A_n$ be a free automaton that
solves $\approxfreq(\epsilon)$  and has kernel $M$. Let $M^e$ be the
smallest dimension subspace of $\R^n$ containing $M$. Let $V_1, V_2$
be a collection of vectors that forms an orthonormal basis for
$\R^n$ such that $V_2$ spans $M^e$ and $V_1$ spans $\R^n/M^e$. Then,
for $ 1/\sqrt{6n} < \epsilon \le \frac{1}{8}$, $\rank(V_1) \ge
\frac{1}{72 \epsilon^2}$.
\end{lemma}
\begin{proof}
\newcommand{\ui}{\ensuremath{u^{(i)}}}
\newcommand{\vi}{\ensuremath{v^{(i)}}}
Since, $V_1$ has orthogonal columns
\begin{align}\label{eq:diag1}\norm{V_1V_1^Te_i}_2^2 = \norm{V_1^T e_i}_2^2 =
(V_1V_1^T e_i)_i \enspace .
\end{align}
Therefore,
\begin{align*} \text{ trace}(V_1V_1^T) =
 \sum_{i=1}^n (V_1V_1^T e_i)_i = \sum_{i=1}^n \norm{V_1 V_1^T
 e_i}_2^2
\end{align*}
The trace of $V_1 V_1^T$ is the sum of the eigenvalues of
$V_1V_1^T$. Since, $V_1$ is orthogonal columns and has rank
$\rank(V_1)$, $V_1V_1^T$  has eigenvalue 1 with multiplicity
$\rank(V_1)$ and eigenvalue 0 with multiplicity $n - \rank(V_1)$.
Thus, $\text{trace}(V_1 V_1^T) = \rank(V_1) = r$ (say). It follows
that
\begin{align} \label{eq:trace1}
r = \text{trace}(V_1V_1^T) = \sum_{i=1}^n \norm{V_1 V_1^T
 e_i}_2^2\enspace .
\end{align}
Further,
\begin{align} \label{eq:trace2}
\sum_{i=1}^n \norm{V_1V_1^Te_i}_1  &\le \sum_{i=1}^n
\norm{V_1V_1^Te_i}_2 \sqrt{n} , ~~~~~~~  \text{ since, $\norm{x}_1
\le \norm{x}_2 \sqrt{n}$} \notag\\ &\le \sqrt{n} \left(\sum_{i=1}^n
\norm{V_1V_1^Te_i}_2^2\right)^{1/2} n^{1/2} , ~~~~~ \text{ by
Cauchy-Schwartz
inequality} \notag \\
&  = n \sqrt{k} ~~~~~ \text{ by ~\eqref{eq:trace1}}\enspace .
\end{align}
Let  \begin{align*} J  & = \{V_1 V_1^Te_i \mid 1\le i \le n \text{
and } \norm{V_1V_1^Te_i}_2^2  \le 3r/n\}, \text{ and }\\
 K  & = \{V_1
V_1^Te_i \mid 1 \le i \le n \text{ and } \norm{V_1V_1^T e_i}_1 \le 3
\sqrt{r} \}  \enspace . \end{align*} Therefore, by
~\eqref{eq:trace1} and ~\eqref{eq:trace2}, $$\abs{J} \ge
\frac{2n}{3}  \text{ and } \abs{K} \ge \frac{2n}{3} \enspace .
$$
Hence, $J \cap K \ne \phi$, that is, there exists $i$ such that
$\norm{V_1 V_1^T e_i}_2 \le (3r/n)^{1/2}$ and $\norm{V_1 V_1^T
e_i}_1 \le 3 \sqrt{r}$. Since, $e_i - V_1V_1^Te_i  = V_2V_2^Te_i \in
M^e$, therefore,
\begin{align*}
\epsilon \ge \error(e_i-V_1V_1^Te_i, 0) = \frac{\norm{e_i-
V_1V_1^T}_{\infty}}{\norm{e_i - V_1V_1^T}_1} \enspace .
\end{align*}
Therefore, \begin{align} \label{eq:trace3} \norm{e_i -
V_1V_1^Te_i}_{\infty} & \le \epsilon \norm{V_1V_1^Te_i-e_i}_1
\enspace .\end{align} By ~\eqref{eq:diag1},
$$(V_1 V_1^T e_i)_i = \norm{V_1V_1^Te_i}_2^2 \le \frac{3r}{n} \enspace
.$$ Therefore,
$$\norm{e_i - V_1V_1^Te_i}_{\infty} \ge \abs{(e_i - V_1V_1^Te_i)_i}
= 1- \norm{V_1V_1^Te_i}_2^2 \ge 1 - \frac{3r}{n}, ~~\text{ by
~\eqref{eq:diag1} and since $V_1V_1^Te_i \in J$} \enspace .
$$
Substituting in ~\eqref{eq:trace3},
\begin{align*}
1- \frac{3r}{n}  \le \norm{e_i - V_1V_1^Te_i}_{\infty}  \le \epsilon
\norm{V_1V_1^Te_i - e_i}_1  & \le \epsilon\bigl(\norm{V_1V_1^T e_i}
+1\bigr),
\text{ by triangle inequality}  \\
& \le \epsilon (3 \sqrt{r}+1), ~~\text{ since, $V_1V_1^T e_i \in K$
\enspace .}
\end{align*}
Simplifying, $ r \ge  \min\left(n/6, 1/(36 \epsilon^2) - 1/9
\epsilon) \right)$. Therefore, for $ 1/\sqrt{6n} < \epsilon \le
\frac{1}{8}$, $r \ge \frac{1}{72 \epsilon^2}$. \qed
\end{proof}

\begin{lemma} \label{lem:free} Let $\frac{1}{6\sqrt{n}} \leq
\epsilon < \frac{1}{8}$. Suppose  $A_n$ be a free  automaton that
uses $s(n,m)$ bits on the work-tape
 to solve $\approxfreq(\epsilon)$. Then, $s(n,m)  = \Omega\left(\frac{\log m}{\epsilon^2}  \right)$.
\end{lemma}
\begin{proof} Let $M=$  kernel of $A_n$.
By Lemma~\ref{lem:J},  $\rank(V_1) = n - \dim M^e = \Omega\left(
\frac{1}{\epsilon^2 }\right)$. By Lemma~\ref{lem:numstates2},
$s(n,m) = \Omega((n-\dim M) \log m )$. Since,  $\dim M = \dim M^e$,
the result follows. \qed
\end{proof}

\subsection{General path independent automata} \label{sec:pi2}
We now show that for the problem $\approxfreq(\epsilon)$, it is
sufficient to consider  free automata. Let $A_n$ be a
path-independent automaton that  solves $\approxfreq(\epsilon)$ and
has kernel $M$. Suppose that  $\Z^n/M$ is not  free. Let $M'$ be the
module that removes the torsion from $\Z^n/M$, that is,
\begin{align} \label{eq:Mprime} M' = \{x \in \Z^n \mid  \exists a \in \Z, a \neq 0
\text{ and } ax \in M \}\enspace .
\end{align}

\begin{lemma} \label{lem:modulefree}
 $\Z^n/M'$ is torsion-free.
 \end{lemma}

  \begin{proof}[Of Lemma~\ref{lem:modulefree}.] Suppose $\bar{y} = y+M' $
  is a torsion element in $\Z^n/M'$. Then, there exists $b \in \Z$ and $b \ne 0$ such
that $b\bar{y} = by + M' \in M'$ or that $by \in M'$. Therefore,
there exists $a \in \Z$, $a \neq 0$, such that $by = ax$, for some
$x \in M$, or that, $y = (b^{-1} a) x$ with $b^{-1}a \neq 0$.
Therefore, $y \in M$. Hence, $\Z^n/M'$ is torsion-free. \qed
\end{proof}

\begin{fact} \label{lem:Me} Let $b_1, b_2, \ldots, b_r$ be a basis of $M'$. Then,
$\exists$  $\alpha_1, \ldots, \alpha_r \in \Z -\{0\}$ such that
$\alpha_1 b_1, \ldots, \alpha_r b_r$ is a basis for $M$. Hence, $M^e
= (M')^e$.
\end{fact}

\begin{proof}[Of
Fact~\ref{lem:Me}] It follows from standard algebra that the basis
of $M$ is of the form $\alpha_1 b_1, \ldots, \alpha_r b_r$. It
remains to be shown that the $\alpha_i$'s are non-zero. Suppose that
 $\alpha_1 = 0$. For any $a \in \Z$, $a \neq 0$,
suppose $ax \in M$ and $x \in M'$. Then, $x$ has a unique
representation as $x = \sum_{j=1}^r x_j b_j$. Thus, $ ax =
\sum_{j=1}^r (ax_j) b_j \in M $ and has the same representation in
the basis $\{\alpha_j b_j\}_{j=1,\ldots, n}$. Therefore, $ax_1 = 0$
or $x_1 = 0$ for all $x \in M'$, which is a contradiction.

 Let $\{b_1, b_2, \ldots, b_r\}$ be a basis for
$M'$. Then, by the above paragraph, there exist non-zero elements
$\alpha_1, \ldots, \alpha_r$ such that $\{\alpha_1 b_1, \alpha_2
b_2, \ldots, \alpha_r b_r\}$ is a basis for $M$. Therefore, over
reals, $(b_1, \ldots, b_r) = (\alpha_1b_1, \ldots, \alpha_r b_r)$.
Thus, $M^e = (M')^e$. \qed
\end{proof}
We  show that if a path independent automaton with kernel $M$ can
solve $\approxfreq(\epsilon)$, then a free automaton with kernel $M'
\supset M$ can solve $\approxfreq(4\epsilon)$.
\begin{lemma} \label{lem:pathindependent}
Suppose $A_n$ is a path independent automaton  for solving
\approxfreq$(\epsilon)$  and has kernel $M$. Then, there exists a
free automaton $B_n$ with kernel  $M'$ such that  $M' \supset M$,
$\Z^n/M'$ is free, and $\error(\min_{\ell_1}(x+M'),x) \leq
4\epsilon$ .
\end{lemma}

\begin{proof}[Of Lemma~\ref{lem:pathindependent}]
Let $M$ be the kernel of $A_n$ and let $M'$ be as defined  in
~\eqref{eq:Mprime}, so that $\Z^n/M'$ is free.\eat{ Since $A_n$
solves $\approxfreq(\epsilon)$ and is path-independent with kernel
$M$,
 for each coset $x+M$, there exists $\hat{x} \in \Z^n$ such that
$\error(\hat{x}, x+M) \leq \epsilon$.} For $x\in \Z^n$, define
$h(x+M') = \min_{\ell_1}(x+M')$. \eat{Denote this as $h(x+M')$, that
is,
$$ h(x+M') = \text{argmin}_{y \in x+M'} ~~\norm{y}_1 \enspace . $$
Let $y \in x+M'$.} Let $y \in x+M'$. Then, $y \in x_1 + M$ for some
$x_1$. Let $\hat{y} = \out_{A_n}(x_1+M)$ denote the output of  $A_n$
for an input stream with frequency in  $x_1 +M$ (they all return the
same value, since, $A_n$ is path independent and has kernel $M$) and
let $y'= \min_{\ell_1}(x_1+M)$. Let $h$ denote $ h(x+M')$ and let
$\hat{h} = \out_{A_n}(h+M)$. Therefore,
\begin{align} \label{eq:tsum1}
\error(h,y) & = \frac{\norm{y-h}_{\infty}}{\norm{y}_1}\leq \frac{
\norm{y-\hat{y}}_{\infty}}{\norm{y}_1} + \frac{\norm{\hat{y} -
y'}_{\infty}}{\norm{y}_1}
 + \frac{\norm{y' - h}_{\infty}}{\norm{y}_1}
\end{align}
The first and the  second  terms above are bounded by $\epsilon$
 as follows. The first term $\frac{
\norm{y-\hat{y}}_{\infty}}{\norm{y}_1}  = \error(\hat{y},y) \leq
\epsilon$, since, $y \in x_1 +M$ and $\hat{y}$ is the estimate
returned by $A_n$ for this coset. The second term
$$\frac{\norm{\hat{y} - y'}_{\infty}}{\norm{y}_1}
 \leq  \frac{\norm{\hat{y} - y'}_{\infty}}{\norm{y'}_1} = \error(\hat{y}, y') \leq
 \epsilon$$
since, $\norm{y'}_1 \leq \norm{y}_1$ and $y'$ lies in the coset $x_1
+M$. The third term in ~\eqref{eq:tsum1} can be rewritten as
follows.  By Lemma~\ref{lem:modulefree}, $y' -h \in M'$ and $M'
\subset M^e$. Therefore,
\begin{align*}
\frac{\norm{y' - h}_{\infty}}{\norm{y}_1} &  \leq  \frac{\norm{y' -
h}_{\infty}}{\norm{y'-h}_{1}} \cdot \frac{\norm{y'-h}
_{1}}{\norm{y'}_1}, ~~~~\text{ since, } \norm{y'}_1 \leq
\norm{y}_1 \\
& \leq \epsilon \cdot \frac{\norm{y'}_{1} +
\norm{h}_{1}}{\norm{y'}_1}  ~~~\text{by Lemma~\ref{lem:errorMe} and
by triangle inequality }
\\
& \leq 2 \epsilon , ~~~~\text{ since, $ \norm{h}_1 \leq \norm{y'}_1$
}
\end{align*}
By ~\eqref{eq:tsum1}, $\error(h,y) \leq \epsilon + \epsilon +
2\epsilon =  4\epsilon$. The automaton $B_n$ with kernel $M'$ is
constructed as in Lemma~\ref{lem:makepi}.
 \qed
\end{proof}

\begin{lemma} \label{lem:lbfree}
Suppose $\frac{1}{24\sqrt{n}} \leq \epsilon < \frac{1}{32}$. Let
$A_n$ be a path independent  automaton that solves
$\approxfreq~(\epsilon)$.  If  $A_n$ has kernel $M$, then, \eat{ and
uses space $s(n,m)$ bits on its work-tape. Then, $s(n,m) + \log
\abs{Q} = \Omega((n-\dim M) \log n)$ and} $n-\dim M =
\Omega\left(\frac{1}{\epsilon^2 }\right)$.
\end{lemma}

\begin{proof} By
Lemma~\ref{lem:pathindependent}, there exists a  free automaton
$A'_n$ with kernel $M' \supset M$  that solves
$\approxfreq(4\epsilon)$. Therefore,  $n - \dim M \geq n - \dim M' =
\Omega\left(\frac{1}{\epsilon^2 }\right)$, by
Lemma~\ref{lem:free}.\qed
\end{proof}

\section{Path reversible automata}\label{sec:prpr}
 In this section, we show that given a path reversible automaton
 $A_n$, one can construct a path independent automaton $B_n$ that is
 an output restriction of $A_n$ and $\Space(B_n,m) \le \Space(A_n,m)
 + O(\log n)$.
 Let $A_n$ be a path reversible
automaton.  For $f \in \Z^n$, define $ \phi_{A_n}(f) = \{s \mid
\exists \sigma $ \text{ s.t. } $o \oplus \sigma = s $ and $\trace
\sigma = f \}$. The kernel of $A_n$ is defined as follows: $ M =
M_{A_n} =  \{ f \mid o \in \phi_{A_n}(f) \} $. Let $C = C(A_n)$ be
the set of reachable configurations from the initial state $o$ of
$A_n$ and let $C_m = C_m(A_n)$ denote the subset of $C(A_n)$ that
are reachable from the initial state $o$ on input streams $\sigma $
with $\abs{\sigma} \le m$. Define a binary relation over $C$ as
follows: $s \sim t $ if there exists $f \in \Z^n$ such that $s,t \in
\phi_{A_n}(f)$.
\begin{lemma} \label{lem:prmodule} \begin{enumerate} \item  $M$ is a sub-module of $\Z^n$.
\item If $f-g \in M$ then $\phi_{A_n}(f) = \phi_{A_n}(g)$, and,  if
$\phi_{A_n}(f) \cap \phi_{A_n}(g)$ is non-empty, then, $f-g \in M$.
\item  The relation $\sim$ over $C$ is an equivalence relation. \item  The
map $[s] \mapsto f + M$, for $s \in \phi_{A_n}(f)$, is well-defined,
1-1 and onto. \end{enumerate}
\end{lemma}

\begin{proof} [Of Lemma~\ref{lem:prmodule}, part 1.]
Since the empty stream has frequency 0,  $0 \in M$. Suppose $f \in
M$. There exists $\sigma$ such that $\trace \sigma = f$ and $o
\oplus \sigma = o$. By path reversibility, $ o = o \oplus \sigma
\circ \sigma^r = o \oplus \sigma^r $. Since $\trace \sigma^r = -
\trace \sigma =  -f$, therefore, $-f \in M$. Now suppose $f,g \in
M$. Then there exists $\sigma, \tau$ such that $\trace \sigma = f,
\trace \tau = g$, $o \oplus \sigma = o$ and $o\oplus \tau = o$.
Therefore, $ o \oplus \sigma \circ \tau = o \circ \tau = o$.  Since,
$\trace \sigma \circ \tau = \trace \sigma + \trace \tau = f+g$,
therefore, $f+g \in M$. \qed
\end{proof}
\begin{proof} [Of Lemma~\ref{lem:prmodule}, part 2.]
Suppose $f = g + h$, for some $h \in M$. Then, there exists $\sigma$
such that $o \oplus \sigma = o$ and $\trace \sigma = h$. Let $a \in
\phi_{A_n}(g)$ and let $\tau$ be a stream such that $o \oplus \tau =
a$ and $\trace \tau = g$. Then, $ o \oplus \sigma \oplus \tau = o
\oplus \tau = a$, and $\trace \sigma \oplus \tau = \trace \sigma +
\trace \tau = h + g = f$. Therefore, $a \in \phi_{A_n}(f)$, or,
$\phi_{A_n}(g) \subset \phi_{A_n}(f)$. Reversing the roles of $f$
and $g$, we have, $\phi_{A_n}(f) \subset \phi_{A_n}(g)$, or that,
$\phi_{A_n}(f) = \phi_{A_n}(g)$. This proves the first assertion of
the lemma. Conversely, Suppose $a \in \phi_{A_n}(f) \cap
\phi_{A_n}(g)$. Then, there exist streams $\sigma$ and $\tau$ such
that $\trace \sigma = f$, $\trace\tau = g$ and  $o \oplus \sigma = o
\oplus \tau = a$. By path reversibility, $ a \oplus \tau^r = o$.
Therefore, $ o\oplus \sigma \circ \tau^r = a \circ \tau^r = o$, and
$\trace \sigma \circ \tau^r = \trace \sigma + \trace \tau^r = f -g$.
Therefore, $o \in \phi_{A_n}(f-g)$ and so $f-g \in M$. \qed
\end{proof}
\begin{proof}[Of Lemma~\ref{lem:prmodule}, part 3.]
By definition, $\sim$ is reflexive and symmetric. Suppose that $s
\sim t $ and $t \sim u$. Then, there exists $f,g \in \Z^n$ such that
$s,t \in \phi_{A_n}(f)$ and $t,u \in \phi_{A_n}(g)$. Therefore, $t
\in \phi_{A_n}(f) \cap \phi_{A_n}(g)$. Hence, $f-g \in M$ and so
$\phi_{A_n}(f) = \phi_{A_n}(g)$. Thus,
 $s \sim u$. \qed
\end{proof}
\begin{proof}[Of Lemma~\ref{lem:prmodule}, part 4.]
Suppose $s \in \phi_{A_n}(f) \cap \phi_{A_n}(g)$, then, $f-g \in M$,
by Lemma~\ref{lem:prmodule}, part 2, or that, $f+M = g+M$. Hence,
the map is well-defined. Suppose $[s]$ and $[t]$ both map to $f+M$.
Then, $s,t \in \phi_{A_n}(f)$, and so $s\sim t$ and therefore, $[s]
= [t]$. Hence the map is \emph{1-1}. For $f \in \Z^n$,
$\phi_{A_n}(f)$ is non-empty and  for any $s \in \phi_{A_n}(f)$,
$[s]$ maps to $f+M$, proving ontoness.\qed
\end{proof}
\eat{\begin{lemma}\label{lem:reva}  If $f-g \in M$ then
$\phi_{A_n}(f) = \phi_{A_n}(g)$. If $\phi_{A_n}(f) \cap
\phi_{A_n}(g)$ is non-empty, then, $f-g \in M$.
\end{lemma}

\begin{lemma} \label{lem:prequiv}The relation $\sim$ over $C$ is an equivalence relation.
\end{lemma}
For $s \in C$, let $[s]$ denote the equivalence class to which $s$
belongs.
\begin{lemma} \label{lem:mappr}The map $[s] \mapsto f + M$, for $s \in \phi_{A_n}(f)$, is well-defined, 1-1
and onto.
\end{lemma}}
Let $B_n$ be a path independent  stream automaton whose
configurations are the set of cosets of $M$ and whose transition is
defined as by the sum of the cosets, that is, $f + (x+M) = (f+x)
+M$, constructed using Lemma~\ref{lem:makepi}. Its output on an
input stream $\sigma$ is defined as:
\newcommand{\myoutput}{\text{out}}
\begin{center}
$\out_{B_n}(\sigma) = \text{choice } \{\text{output of $A_n$ in
configuration $s \mid   s \in \phi_{A_n}(\freq
\sigma)$}\}$\end{center} where, $\text{choice } S$ returns some
element from its argument set $S$.
\begin{lemma} \label{lem:freqeq}
$B_n$ is an output restriction  of $A_n$.
\end{lemma}

\begin{proof}
$f+M = g+M$ if and only if $\phi_{A_n}(f) = \phi_{A_n}(g)$.
Therefore, $\myoutput_B(\sigma)$ is well-defined. Further, by
definition of $\myoutput_B$, $\myoutput_B(\sigma) = $ the output of
$A$ in some configuration $s$, where, $s \in \phi_{A_n}(\freq
\sigma)$. Thus, $B_n$ is an output restriction of  $A_n$. \qed
\end{proof}
We can now prove the main lemma of the section.
\begin{lemma} \label{lem:prmain} Let $A_n$ be a  path reversible automaton with kernel $M$\eat{ that
uses $s_A(m,n)$ bits of  work space and has $\abs{Q_A}$ states in
its finite control}. Then, there exists a path independent automaton
$B_n$ with kernel $M$ that is an output restriction of $A_n$ such
that  $\log\abs{C_m(A_n)} +O(\log n) \geq \Space(B_n,m)$, for $m \ge
1$.
\end{lemma}
\begin{proof} Let  $B_n$ be constructed in the manner described above. By Lemma~\ref{lem:freqeq},
is an output-restriction of $A_n$. Since the map $[s] \rightarrow
f+M$, for $s \in \phi_{A_n}(f)$ is 1-1 and onto
(Lemma~\ref{lem:prmodule}, part 4), therefore, for every $m$, each
reachable configuration of $B_n$ after processing streams $\sigma$
with $\freq\sigma \in  [-m\ldots m]^n$ can be associated with a
disjoint aggregate of configurations of $A_n$. The number of
reachable configurations of $B_n$ after processing streams with
frequency in $[-m \ldots m]^n$ is $~\abs{\{x+M \mid x \in [-m\ldots
m]^n\}}$. Thus, $\abs{C(A_n)} \geq \abs{\{x+M \mid x \in [-m\ldots
m]^n\}}$.  By Lemma~\ref{lem:makepi}, $\Space(B_n,m) = \log
~\abs{\{x+M \mid x \in [-m\ldots m]^n\}} +O(\log n)$. Combining, we
obtain the statement of the lemma. \qed
\end{proof}
\paragraph{Remarks.} The above  procedure transforms  a path reversible
automaton $A_n$ to a path-independent automaton $B_n$ such that
$\log\abs{C_m(A_n)} +O(\log n) \geq \Space(B_n,m)$, for all $m \ge
1$. However, the arguments only use the property that the transition
function $\oplus_{A_n}$ is path reversible, and the fact that the
subset of reachable configurations $C_m(A_n)$ on streams of size at
most $m$ is finite. The argument is more general and also applies to
computation performed by an  infinite-state deterministic automaton
in the classical sense that returns an output after it sees the end
of its input, with set of states $C$, initial state $o$ and  a
path-reversible transition function $\oplus'_{A_n}$. The above
argument shows that such an automaton $A_n$  can be simulated by a
path-independent stream automaton $B_n$  with finite control and
additional space overhead of $O(\log n)$ bits, such that $B_n$ is an
output-restriction of $A_n$. We will use this observation in the
next section.

\section{Path non-reversible automata} \label{sec:pnr} In this
section, we show that corresponding to every general stream
automaton $A_n$, there exists a path reversible automaton  $A'_n$
that is an output-restriction of $A'_n$, such that $\Space(A_n,m)
\ge \log \abs{C_m(A'_n)}$.  By Lemma~\ref{lem:prmain}, corresponding
to any path reversible automaton $A'_n$, there exists an
output-restricted and path independent automaton $B_n$, such that
$\log \abs{C_m(A'_n)} \ge \Space(B_n,m) - O(\log n)$. Together, this
proves a basic property of stream automata, namely, that, for every
stream automaton $A_n$, there exists a path-independent stream
automaton $B_n$ that is an output-restriction of $A_n$  and
$\Space(B_n,m) \le \Space(A_n,m) + O(\log n)$. We  construct the
path-reversible automaton $A'_n$ only to the extent  of designing  a
path-reversible transition function $\oplus_{A'_n}$, a set of
configurations $C(A'_n)$ and specifying the output of $A'_n$ if the
end of the stream is met while at any $s \in C(A'_n)$. As per the
remarks at the end of the previous section, this is sufficient to
enable the construction of the path-independent automaton $B_n$ from
$A'_n$.

\subsection{Defining  reversible transition function from
stream automata} \label{app:pnr} In this section, we present
detailed (existential) construction of constructing a reversible
transition function $\oplus' = \oplus_{A'_n} $ from a given general
stream automaton $A_n$ with transition function $\oplus=
\oplus_{A_n}$. Let $C = C(A_n)$ denote the space of configurations
of $A_n$ and let $C_m = C_m(A_n)$ denote the subset of $C(A_n)$ that
are reachable from $o$ on input streams of size at most $m$.

Consider a directed graph $G = (C,E)$ where,  $C = C(A_n)$ is the
set of vertices and  there is a directed edge from $s$ to $t$
provided there is some stream $\sigma$ such that $\trace \sigma = 0$
and $s \oplus \sigma = t$. Define the equivalence relation $s
\sim_{G} t$ if there is a directed path from $s$ to $t$ in $G$ and
vice-versa. Let $[s]_{\sim_G}$ denote the equivalence class to which
a configuration $s$ belongs.  Define the equivalence class
restricted to the vertices of $C_m$ as $[s]_{\sim_{G_m}} =
[s]_{\sim_G}\cap C_m$. An equivalence class $[s]_{\sim_{G_m}}$  that
satisfies the property that for every stream $\sigma$ with $\trace
\sigma = 0$ and $s \oplus \sigma \in C_m$, we have $s \oplus \sigma
\in [s]_{\sim_{G_m}}$, are called \emph{terminal} equivalence
classes.
\begin{lemma} \label{lem:terminal} For every $m \geq 1$ and  $u \in C_m$,
there exists $s = s(u) $ reachable from $u$ in $G_m$ such that
$[s]_{\sim_{G_m}}$ is a terminal equivalence class.
\end{lemma}

\begin{proof}[Of Lemma~\ref{lem:terminal}.] Let $u_0$ be a vertex reachable from $u$ in $G_m$.
If $[u_0]_{\sim_{G_m}}$ satisfies the property stated in the lemma,
then, we are done. Otherwise, there exists $\sigma$ such that
$\trace \sigma = 0$ and $u_1 = u_0 \oplus \sigma \in C_m - [u_0]$.
 We now
 iteratively construct the sequence $[u_1]_{\sim_{G_m}}, [u_2]_{\sim_{G_m}}, \ldots,
 $ in this manner.
Suppose that two equivalence classes in this sequence are the same,
that is, suppose $[u_i]_{\sim_{G_m}} = [u_j]_{\sim_{G_m}}$. Then,
there exists a directed path from $u_i$ to $u_j$ and vice-versa and
therefore, $[u_i]_{\sim_{G_m}} = \ldots =  [u_j]_{\sim_{G_m}}$, that
is, the iteration terminates. Since, $C_m$  is finite, the iterated
sequence of equivalence classes of $\sim_{G_m}$ terminates. The last
equivalence class of this sequence satisfies the property of the
lemma. \qed
\end{proof}

Define the mapping $\alpha_m: C_m \rightarrow C_m$ as follows:
$\alpha_m(s) = $ some member  of some  terminal equivalence class
reachable from  $s$ (for e.g.,  the member with least lexicographic
value among all candidates). Fix $s \in C$ and consider the sequence
$\{\alpha_m(s)\}_{m \geq 1}$. If this sequence is finite, then, one
can define $\alpha(s)$ to be a final element of the sequence.
Otherwise, we use a standard technique of passing to the infinite
case by associating $s$ with `consistent' infinite sequences
$\bar{s} = \{\alpha_m(s)\}_{m \geq 1}$. \eat{ If $\bar{s}$ is
finite, then the modified transition function $\oplus'_n$ is defined
as follows. The infinite case and the proof of Lemma~\ref{lem:nrrev}
for the cases when $\{\alpha_m(s)\}_{m \ge 1}$ is  infinite is
presented below.
\begin{align*}
\alpha(s) \oplus' e_i = \alpha(\alpha(s) \oplus e_i)  \text{ and }
\alpha(s) \oplus' -e_i = \alpha(\alpha(s) \oplus -e_i), i=1,2,
\ldots, n\enspace .
\end{align*}

\begin{proof} [Of Lemma~\ref{lem:nrrev} ($\bar{s}$ finite.)] We show the proof for the case when the sequence
$\{\alpha_m(s)\}$ is finite; the proof for the infinite case is
given below.  Let $\alpha(s) = t$. Suppose $t\oplus e_i = u_1$ and
$\alpha(u_1) = u_2$, where, $[u_2]$ is a terminal class. Therefore,
there is a stream $\sigma$ with $\freq \sigma = 0$ such that $u_1
\oplus \sigma = u_2$. Further, $t \oplus e_i \circ \sigma \circ -e_i
\in [t]$, since, $\trace e_i \circ \sigma \circ -e_i = 0$.
Therefore, $u_2 \oplus -e_i \in [t]$. Since $[t]$ is a terminal
equivalence class, $\alpha(u_2 \circ -e_i) = t$.\qed
\end{proof}
}

\begin{lemma} \label{lem:nrrev} For  $s \in C$, $\alpha(s) \oplus'
e_i \circ -e_i = \alpha(s)$ and $\alpha(s) \oplus' -e_i \circ e_i =
\alpha(s)$.
\end{lemma}

\begin{proof}[Of Lemma~\ref{lem:nrrev} ]
A configuration  $s$ is first identified with the infinite sequence,
$\bar{s} = \{\alpha_m(s)\}_{m \geq 1}$. Recall that the definition
of $\alpha_m(s)$ allows flexibility in the choice of a terminal
class of $\sim_{G_m}$. We now ensure that the choices are made in a
consistent manner as follows. For each $m$, there is a path $P_m(s)$
from $s$ to a vertex in the equivalence class $\alpha_m(s)$. By
consistent choices across $m$, we mean that the $P_{m+j}(s)$ is an
extension of the path $P_m(s)$, for each $j > 0$, and for each $s
\in C$. From now

The transition function $\oplus'$ is defined in two steps. First, we
define an intermediate function $\oplus_1$.
\begin{align} \label{eq:oplus1}
\bar{s} \oplus_1 e_i = \{ \alpha_m(\alpha_m(s) \oplus e_i) \}_{m
\geq 1}
\end{align}
Sequences are allowed to have the undefined element $\bot$, since,
it is possible that $s \not\in C_m$ and hence $\alpha_m(s)$ is not
defined. However, if $\alpha_m(s) $ is defined, then,
$\alpha_{m+j}(s)$ is defined, for all $j > 0$. This implies that the
undefined elements, if they occur,  form a prefix of the sequence
$\bar{s}$.

We now attempt to  prove Lemma~\ref{lem:nrrev} for the transition
function $\oplus_1$. Let $m_0$ be the smallest $m$ for which
$\alpha_m(s) \oplus e_i$ is well-defined. Then, for all $m \ge m_0$,
both $\alpha_m(s) \oplus e_i$  and $\alpha(\alpha_m(s) \oplus
e_i)\oplus -e_i $ are also well-defined. The arguments in the finite
case of Lemma~\ref{lem:nrrev} hold for each member  $m \geq m_0$.
The same can be said for $\alpha_m(s) \oplus -e_i$. Thus, the two
sequences
$$\{\alpha_m(s)\}_{m \ge 1} \text{ and } \{\alpha_m(\alpha_m(\alpha_m(s) \oplus
e_i) \oplus -e_i)\}_{m \ge 1}
$$
differ at most in a finite prefix, where, the \emph{RHS} sequence
may have more $\bot$ elements than the sequence on the \emph{LHS}.

To resolve this problem, we define a relation $\cong$ between pairs
of infinite sequences.
\begin{align*}
\{u_m\}_{m \geq 1} \cong \{v_m\}_{m \geq 1} \text{ if  $u_m$ and
$v_m$ differ in a finite initial prefix. }
\end{align*}
A finite sequence $u_1, \ldots, u_r$ is modeled as an infinite
sequence $u_1, \ldots, u_r, u_r, u_r, \ldots $ whose last term is
repeated. It is straightforward to see that $\cong$ is an
equivalence relation on the family of sequences. It now follows that
$$\{\alpha_m(s)\}_{m \ge 1} \cong \{\alpha_m(\alpha_m(\alpha_m(s) \oplus
e_i) \oplus -e_i)\}_{m \ge 1} \enspace .
$$
For each configuration $s$ in the original automaton, we associate
it with  $[s]_{\cong}$ as follows.
\begin{align*}
[s]_{\cong} \stackrel{\text{def}}{=} [ ~\{\alpha_m(s)\}_{m \geq
1}~]_{\cong}
\end{align*}
The transition function $\oplus'$ is now defined as follows.
\begin{align*}
[s]_{\cong} \oplus e_i  & = [  ~\{\alpha(\alpha_m(s) \oplus e_i)
\}_{m
\geq 1} ]_{\cong}  \text{ and } \\
[s]_{\cong} \oplus -e_i & = [~ \{\alpha(\alpha_m(s) \oplus -e_i)
\}_{m \geq 1} ]_{\cong}
\end{align*}
 It now follows, by repeating the arguments in the previous
paragraph, that
\begin{align*}
[s]_{\cong} \oplus' e_i \circ -e_i = [s]_{\cong} \enspace .
\end{align*}
This proves Lemma~\ref{lem:nrrev}, with $\alpha(s)$ defined as $
[~\{\alpha_m(s)\}_{m \geq 1} ]_{\cong}$. \qed
\end{proof}
The map $s \mapsto \alpha(s)$ maps $s$ to a congruence class over
the space of consistent infinite sequences. Define $C'_m =
\{\beta(s) \mid s \in C_m\}$. Therefore, $\abs{C'_m} \leq \abs{C_m}$
for all $m \geq 1$.

A path reversible automaton $A'_n$ is defined as follows. Initially
$A'_n$ is in the state $\alpha(o)$. After reading a stream record
(one of the $e_i$'s or $-e_i$'s), $A'_n$ uses the transition
function $\oplus'$ instead of $\oplus$ to process its input.
However, $s \oplus' \sigma = \alpha(s \oplus \sigma)$, where,
$\alpha(t)$ is  a set  (possibly infinite) of states that cause
$A_n$ to transit from configuration  $t$ on some input $\sigma'$,
with $\freq \sigma' = 0$.  \emph{Equivalently, this can be
interpreted as if $\sigma'$ has been inserted into the input tape
just after $A_n$ reaches the configuration $s$ and before it
processes the next symbol--hence, $A'_n$ is an output-restriction of
$A_n$ and is equally correct for frequency-dependent computations.
This is the main idea of this construction.} Thus, transitions of
$\oplus'$ are equivalent to inserting some specifically chosen
strings $\sigma_1, \sigma_2, \ldots$, each having $\freq =0$, after
reading each letter (i.e., $\pm e_i$) of the input. The output of
$A'_n$ on input stream $\sigma$ is identical to the output of $A_n$
on the stream $\sigma'$, where, $\sigma'$ is obtained by inserting
zero frequency sub-streams into it. Therefore, $\freq(\sigma') =
\freq(\sigma)$ and $A'_n$ is an output restriction of $A_n$.  By
Lemma~\ref{lem:nrrev}, the transition function  $\oplus'$ is path
reversible. Let $C'=C(A'_n)$ and $C'_m = C_m(A'_n)$.  Since,
$\alpha(s)$ is an equivalence class over $C(A_n)$, the map $s
\mapsto \alpha(s)$ implies that $\abs{C'_m} =\abs{ \{\alpha(s) \mid
s \in C_m\}} \leq \abs{ C_m}$.  Starting from $A'_n$, one can
construct a path independent automaton $B_n$ as per the discussion
in Section~\ref{sec:prpr}.  The arguments in this section do not
show that the transition function $\oplus'$ can indeed by realized
by a Turing machine that has only finite control. This  is
sufficient however, since, the path reversibility of $\oplus'$ is
only used to allow the techniques of Section~\ref{sec:prpr} to be
applicable, and hence to be able  to construct a coset-based path
independent automaton. Since any coset based automaton can  be
realized using
 finite number of states in its finite control (Lemma~\ref{lem:numstates2},
therefore, the final path-independent transition function is
actually a stream automaton $B_n$.) Theorem ~\ref{thm:straut}
summarizes this discussion.
\begin{theorem}[Basic property of computations using  stream automata] \label{thm:straut}
For every stream automaton $A_n$, there exists a path-independent
stream automaton $B_n$ that is an output-restriction of $A_n$ and
$\Space(B_n,m) \le \Space(A_n,m) + O(\log n)$.
\end{theorem}
\begin{proof} Let $\oplus'$ be the transition
function of the  path-reversible automaton
 constructed as described above and let  $B_n$ be the
path-independent automaton obtained by translating $\oplus'$ using
the procedure of Section~\ref{sec:prpr}. Let $C_m$  and $C'_m$
denote the number of reachable configurations of $A_n$ and $A'_n$,
respectively,  over streams with frequency  vector in  $[-m\ldots
m]^n$. Let $s_A = s_A(n,m)$. Let $M$ be the kernel of $B_n$. Then,
$$\abs{Q_A}s_A 2^{s_A} \geq \abs{C_m} \geq \abs{C'_m} \geq
\abs{\{x+M \mid x \in [-m\ldots m]^n\}} \geq (2m+1)^{n - \dim M}
  $$ where, the last two inequalities follow from
Lemma~\ref{lem:prmain}.  Taking logarithms, $ \Space(A_n,m) \geq
\log \abs{\{x+M \mid x \in [-m\ldots m]^n\}} \geq \Space(B_n,m) -
O(\log n)$, by   Lemma~\ref{lem:makepi}. \qed
\end{proof}

\begin{theorem}[\bf Lower bound for $\approxfreq(\epsilon)$]\label{lem:lb}
Suppose that $\frac{1}{24\sqrt{n}} \leq \epsilon < \frac{1}{32} $
and let $A_n$ be a stream automaton that solves
$\approxfreq(\epsilon)$. Then, $\Space(A_n,m)
 = \Omega\left( \frac{\log m }{\epsilon^2 }\right) - O(\log n)$.
\end{theorem}
\begin{proof} By Theorem~\ref{thm:straut}, there exists a
path independent automaton $B_n$ that is an output-restriction of
$A_n$ and $\Space(A_n,m) \ge \Space(B_n,m) -O(\log n)$. By
Lemma~\ref{lem:outrestr}, $B_n$ solves $\approxfreq(\epsilon)$. If
$M$ is the kernel of $B_n$, then by Lemma~\ref{lem:numstates2},
$\Space(B_n) = \Omega((n- \dim M)(\log (2m+1))$.  By
Lemma~\ref{lem:lbfree}, $n-\dim M = \Omega\left( \epsilon^{-2}
\right)$. Thus, \\$~~~\Space(A_n,m) = \Omega((n-\dim M) \log m)
-O(\log n) =
 \Omega\left( \frac{\log m}{\epsilon^2 }\right)- O(\log n) \enspace .
~~~~~~~~~\text{
 \qed } $
\end{proof}
Since, any path-independent automaton is arbitrarily mergeable (see
text  before Lemma~\ref{lem:makepi}), Theorem~\ref{thm:straut}
implies that for any stream automaton $A_n$, there exists an
output-restricted  automaton $B_n$ such that $\Space(B_n, m) \le
\Space(A_n,m) + O(\log n)$, and  the state of $B$ is arbitrarily
mergeable, establishing the claim made in Section~\ref{sec:intro}.

\end{document}